\documentclass[a4paper,conference]{IEEEtran}

\IEEEoverridecommandlockouts
\usepackage{cite}
\usepackage{amsmath,amssymb,amsfonts}
\usepackage{balance}
\usepackage{subfigure}
\usepackage{algorithmic}
\usepackage{graphicx}
\usepackage{textcomp}
\usepackage{xcolor}
\def\BibTeX{{\rm B\kern-.05em{\sc i\kern-.025em b}\kern-.08em
    T\kern-.1667em\lower.7ex\hbox{E}\kern-.125emX}}
\begin{document}

\title{Collecting Channel State Information in Wi-Fi Access Points for IoT Forensics}

\author{

    \IEEEauthorblockN{Fabio Palmese}
    \IEEEauthorblockA{
        \textit{DEIB, Politecnico di Milano} \\
        Milan, Italy\\
        fabio.palmese@polimi.it}
    \and

    \IEEEauthorblockN{Alessandro E. C. Redondi}
    \IEEEauthorblockA{
        \textit{DEIB, Politecnico di Milano} \\
        Milan, Italy\\
        alessandroenrico.redondi@polimi.it
        }

}

\maketitle

\begin{abstract}
The Internet of Things (IoT) has boomed in recent years, with an ever-growing number of connected devices and a corresponding exponential increase in network traffic. As a result, IoT devices have become potential witnesses of the surrounding environment and people living in it, creating a vast new source of forensic evidence. To address this need, a new field called IoT Forensics has emerged. In this paper, we present \textit{CSI Sniffer}, a tool that integrates the collection and management of Channel State Information (CSI) in Wi-Fi Access Points. CSI is a physical layer indicator that enables human sensing, including occupancy monitoring and activity recognition. After a description of the tool architecture and implementation, we demonstrate its capabilities through two application scenarios that use binary classification techniques to classify user behavior based on CSI features extracted from IoT traffic.
Our results show that the proposed tool can enhance the capabilities of forensic investigations by providing additional sources of evidence. Wi-Fi Access Points integrated with \textit{CSI Sniffer} can be used by ISP or network managers to facilitate the collection of information from IoT devices and the surrounding environment. We conclude the work by analyzing the storage requirements of CSI sample collection and discussing the impact of lossy compression techniques on classification performance.

\end{abstract}


\section{Introduction}
\label{sec:introduction}
In the last years, IoT companies are continuously launching new products on the market, as well as adding novel functionalities to the already available ones. There is therefore an exponentially increasing number of connected devices, with a corresponding unprecedented growth of the produced network traffic. The communication activities of such devices introduce a huge amount of potential information that could be leveraged as a source of evidence in forensic investigations: indeed, smart devices become potential witnesses of the surrounding environment and the actions of people living around them. 
For this reason, a new branch of digital forensics, named IoT forensics, recently emerged with the goal of identifying and extracting forensic information from this specific category of devices and from the network traffic they exchange with the Internet and among themselves. 

IoT forensics based on network traffic generally leverages groups of packets captured at the IP or upper layers to perform analysis tasks such as device identification, activity recognition or to identify possible attacks or device misuse.
In this paper, we take a different perspective and focus on possible IoT forensic analysis performed starting from the physical characteristics of the network traffic produced by IoT devices. Motivated by the widespread availability of Wi-Fi IoT consumer devices, we focus here on Channel State Information (CSI) features that can be extracted by (almost) any IEEE 802.11-compliant frame. 
CSI is a fine-grained physical layer indicator that describes the radio channel existing between the transmitter and receiver of each frame, furthermore being able to capture multi-path effects. Due to its successful use as a building block for analysis tasks in the area of wireless sensing (human activity recognition \cite{carm}, gesture recognition \cite{WiFinger}, crowd counting\cite{crowd_counting}, etc.), the CSI indicator is very promising as a tool to enable forensic analysis based on Wi-Fi IoT traffic.\\
%
The extraction of CSI features requires ad-hoc hardware/firmware tools which limit its use to very specific use cases or controlled scenarios. The goal of this work is to present CSI Sniffer, a tool that overcomes the limitations of currently available CSI extraction frameworks by enabling and facilitating the collection and management of CSI features directly from any consumer Wi-Fi Access Point. In detail, the tool integrates one of the existing solutions to collect CSI, namely the Nexmon CSI extractor \cite{nexmon:project}, with the popular OpenWrt operating system for 802.11 access points.
In detail, the contributions of this work are the following:
\begin{itemize}
    \item 
    We present CSI Sniffer in detail, focusing on its implementation and configuration parameters, furthermore presenting its user-friendly features integrated into the OpenWrt Graphical User Interface. 
    \item To demonstrate the capabilities of CSI Sniffer we set up two different scenarios where it may be used as an IoT forensic tool: (i) people occupancy detection in a room and (ii) passage detection through a door. In both cases, CSI Sniffer extracts data from traffic exchanged by consumer IoT devices already present in the environment in an opportunistic fashion, providing the basis for subsequent forensic analysis tasks.  
    \item 
    Forensic analyses are generally performed a-posteriori and require network traffic features to be stored for long periods. As an additional contribution, we also comment on the storage/accuracy trade-off existing in CSI-based forensic analysis tasks providing design guidelines to optimize the storage space required.
\end{itemize}



The rest of the paper is structured as follows: Section \ref{sec:background} provides a background on Wi-Fi sensing focusing on the CSI, Section \ref{sec:project} describes the proposed tool with its implementation details. Section \ref{sec:application} shows possible use cases of the CSI for IoT forensic tasks, with the corresponding experimental results. Section \ref{sec:storage} provides an accurate analysis of the dataset storage occupation. A discussion on the related work is reported in Section \ref{sec:related} while Section \ref{sec:conclusion} concludes the work with some final remarks and future research directions.

\section{Background on CSI extraction}
\label{sec:background}
Extracting physical-layer characteristics from network traffic is very promising for IoT forensics analyses aimed at revealing human activities, since (i) most of the consumer IoT devices use radio communication technologies (e.g., Wi-Fi) and (ii) radio signals are greatly affected by human presence and movements. Two main low-cost physical layer indicators are available: the Received Signal Strength Indicator (RSSI) and the Channel State Information (CSI). The former merely reports the signal power of transmission at the receiver, and it is generally used as a starting point for localization systems \cite{rssi-localization}. The latter constitutes a much more detailed measure of the channel characteristics between the transmitter and the receiver. In detail, CSI represents how wireless signals propagate at certain carrier frequencies along multiple paths\cite{10.1145/3310194}. For a MIMO WiFi system, CSI is a 3D matrix of complex values representing the amplitude attenuation and phase shift of multi-path WiFi OFDM subcarriers. The estimation of the CSI matrix is performed at the receiver side, exploiting the knowledge of the Legacy Long Training Field (L-LTF), inserted in every frame by the transmitter. The L-LTF contains predefined sequences for every OFDM subcarrier: therefore the receiver can estimate the CSI matrix by comparing the received signal with the content of such field \cite{8794643}. 
While the RSSI can be easily estimated by every Wi-Fi Network Interface Card (NIC) and exposed to the user space, the same is not true for CSI measurements. Indeed, there are very few NICs and corresponding firmware that support CSI estimation and make it available to the user for further processing. To the best of our knowledge, the available solutions are limited to the following choices:
\begin{itemize}
    \item Linux 802.11n CSI Tool: allows computing CSI from the Intel Wi-Fi Wireless Link 5300 NIC, which is currently discontinued \cite{linux_csi}.
        \item Atheros CSI Tool: Open source Linux kernel driver supporting Atheros 802.11n PCI/PCI-E chips. Limited to the Atheros Wi-Fi cards but compatible with many operating systems, including embedded devices support (e.g. OpenWrt, Linino) \cite{atheros_tool}
        \item ESP32 CSI Toolkit/ Wi-ESP: Available for the ESP 32 microcontroller. Low-cost solution but with limited processing capabilities. \cite{esp32-csi,atif2020wi}
        \item Nexmon CSI tool: Available for several Broadcom-based smartphones and for all the Raspberry Pi models \cite{nexmon:project}. We use this solution as starting point for our framework.
        
\end{itemize}

\section{Proposed system}
\label{sec:project}

\begin{figure}[t]
	\centering
	\includegraphics[width=\columnwidth]{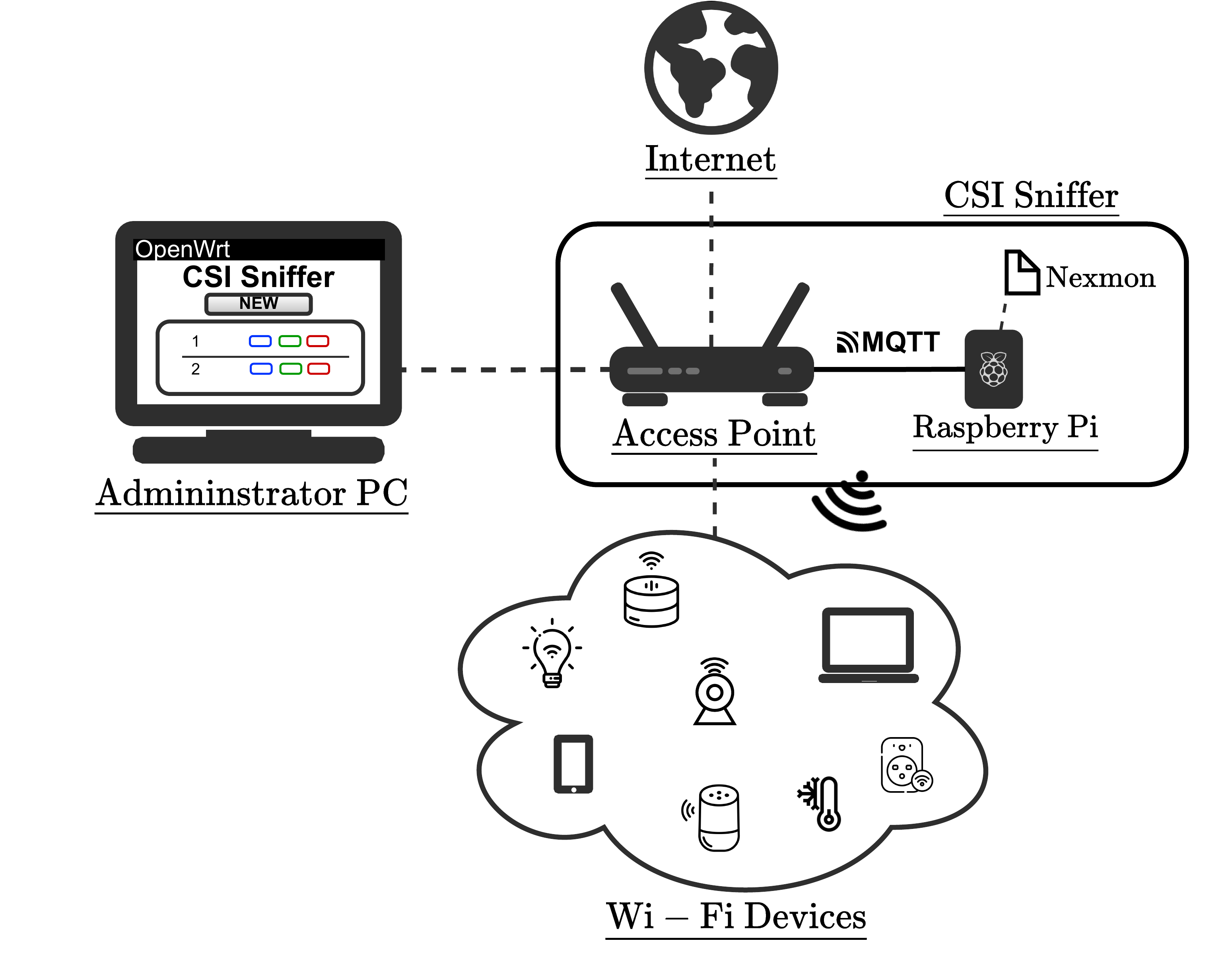}
	\caption{Sketch of the CSI Sniffer architecture}
	\label{fig:csi_architecture}
\end{figure}
This section presents CSI Sniffer, a tool that can be installed in any Wi-Fi Access Point (AP) supporting the OpenWrt firmware that allows users to capture CSI data directly through the AP web interface. This enables gathering CSI data in a user-friendly fashion, facilitating subsequent forensic analysis tasks.

\begin{figure*}[t]
	\centering
	\includegraphics[width=1.8\columnwidth]{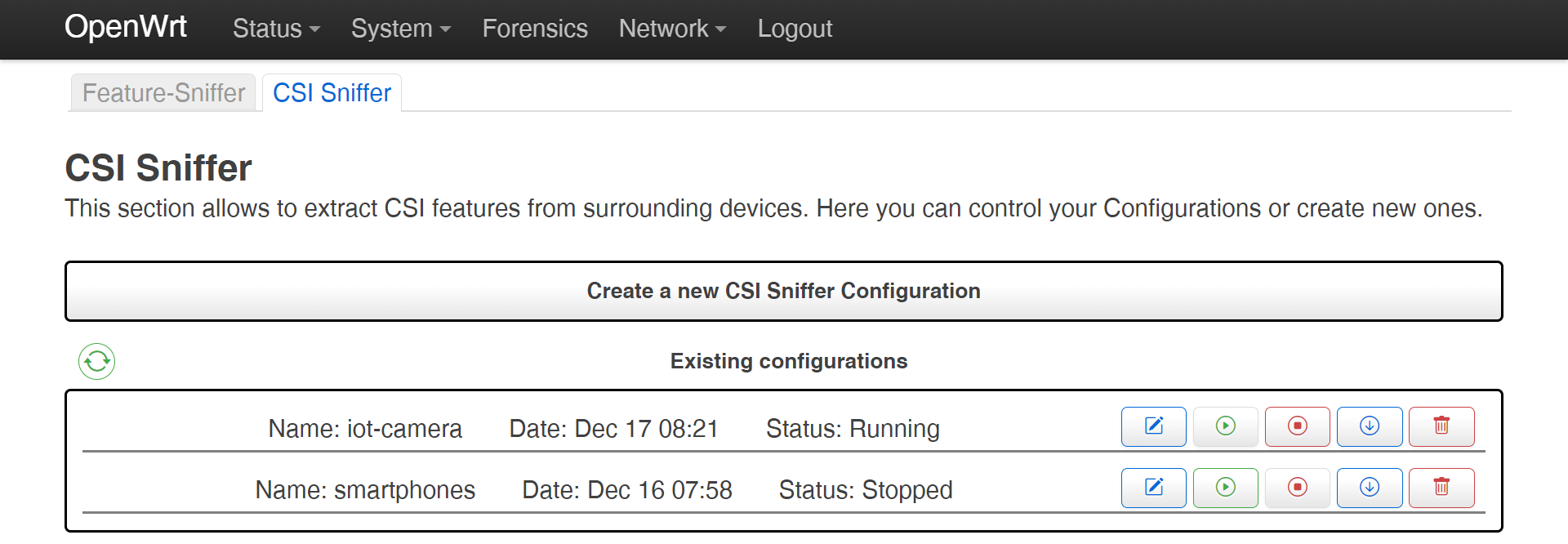}
	\caption{CSI Sniffer User Interface: The user can easily control existing configurations or create new ones}
	\label{fig:csi_home}
\end{figure*}

\subsection{Architecture}
The CSI Sniffer architecture relies mainly on two components: an Access Point as the central entity, and a Raspberry Pi as the CSI collector. The latter is connected to the AP as a client through a standard Ethernet cable.
The Access Point is equipped with the OpenWrt operating system and the LuCI graphical web interface to easily manage the AP functionalities, while the Raspberry Pi is equipped with the Nexmon CSI extractor. The latter allows the computation of CSI features from sniffed Wi-Fi frames, allowing a user to control several aspects of the process (e.g., Wi-Fi sniffing channel, device filter, etc.). The output of Nexmon CSI is a Comma Separated Value (CSV) file where each row corresponds to a captured frame and contains the reception timestamp of the frame, the transmitter MAC address, and the CSI for each OFDM subcarrier, represented as a complex number. A sketch of the proposed architecture is reported in Figure \ref{fig:csi_architecture}.

\subsection{User interface}
The CSI Feature Sniffer user interface is an extension of the LuCI web interface and enhances the functionalities of our previous work \cite{feature-sniffer}. As reported in Figure \ref{fig:csi_home}, the user interface allows creating new \textit{configurations}, each one characterized by different operational parameters. In particular, each configuration is described by:
\begin{itemize}
    \item Name and Description: a unique string used as an identifier and a brief text to use as a description
    \item Frequency Band: the Wi-Fi band the wireless card will use for the sniffing process (2.4GHz or 5GHz).
    \item Channel Bandwidth: the Wi-Fi Channel width. Can be set to 20 and 40MHz for 2.4GHz, and 40 or 80MHz for the 5GHz band.
    \item Channel: the Wi-Fi Channel to listen to.
    \item Device Filter: a list containing the MAC address of the devices to consider in the CSI collection. If left empty, all the devices are considered in the collection phase.
\end{itemize}
Once created through the proper button, each configuration can be controlled with specific buttons to \textit{show/edit} its settings, \textit{start} and \textit{stop} the CSI capture, \textit{download} the resulting output as a CSV file, or \textit{delete} the configuration.

\subsection{Implementation details}
As aforementioned, the AP and the Raspberry PI are connected via Ethernet. Communication between the two devices is performed via the MQTT protocol, chosen for its simplicity and low impact on computational performance. 
Upon booting up the Raspberry PI a specific script is executed to (i) launch an instance of the Mosquitto MQTT broker \cite{mosquitto} and (ii) subscribe the Raspberry PI to the \texttt{start, stop} and \texttt{download} MQTT topics. 
The AP runs a simple MQTT client, which connects to the broker on the Raspberry at startup. Upon clicking the start button on a specific CSI sniffer configuration on the Access Point, an MQTT publish message containing the configuration parameters as a payload (name, devices list, band/channel) is transmitted to the MQTT broker. As the MQTT subscriber on the Raspberry PI receives the configuration parameters, it tunes the Nexmon CSI tool with the corresponding settings and starts the CSI collection. A similar procedure is used to interrupt the process, by publishing a message to the \texttt{stop} topic indicating the name of the configuration to stop. When the user requests the output using the \textit{download} button, a publish message is sent to the \texttt{download} topic, causing the Raspberry PI to prepare the CSI measurements to be sent back as a .CSV file. Operatively, the CSI file is first published from the Raspberry PI to the AP on the \texttt{output} topic and then forwarded to the user space in the payload of the HTTP response to the request triggered by the press of the \textit{download} button.

\section{Application Cases}
\label{sec:application}

\begin{figure*}[t]
	\centering
	\includegraphics[width=1.7\columnwidth]{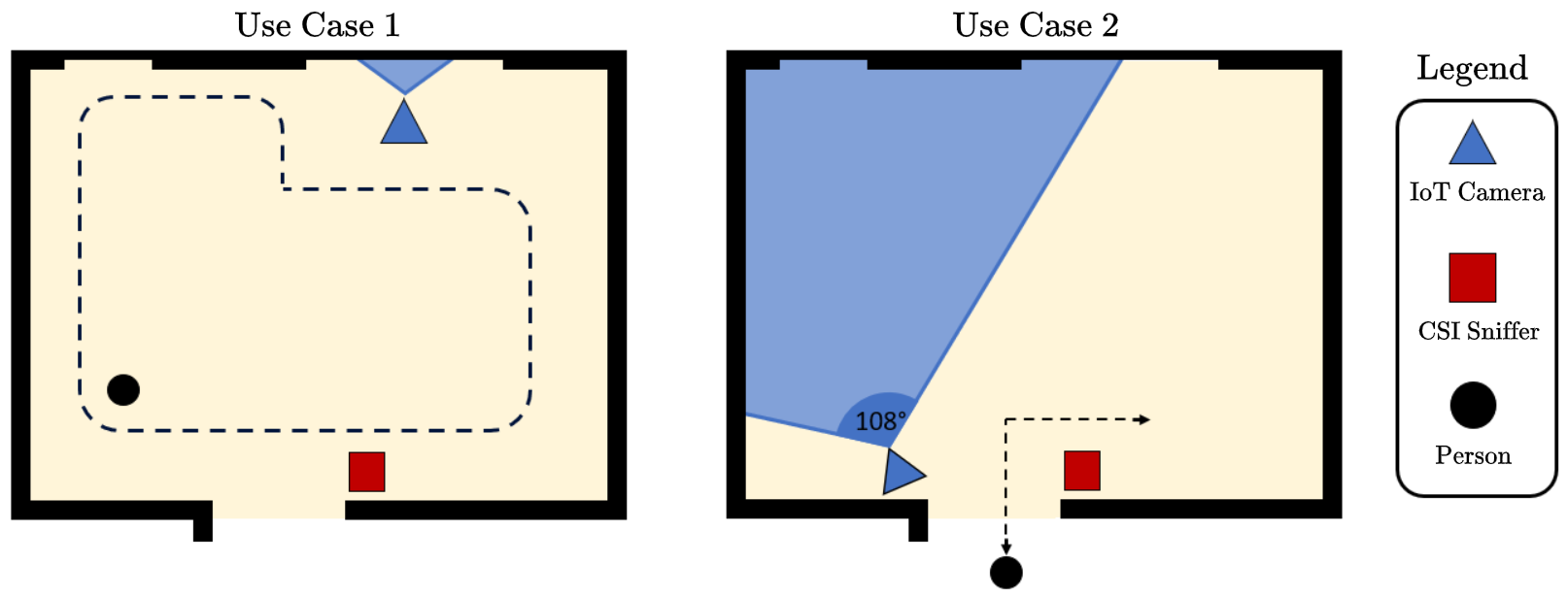}
	\caption{Sketch of the disposition of devices in the room for Use Case 1 (left) and Use Case 2 (right)}
	\label{fig:scenarios}
\end{figure*}

To showcase the potential of CSI Sniffer in realistic IoT forensic scenarios, we set up two different experiments where consumer IoT devices produce network traffic which is opportunistically leveraged to extract CSI measurements and analyze them. We opt for a Teckin TC100 smart camera installed and active in the test environment: note that this choice is driven by the high packet generation rate typical of smart cameras, rather than by the device type itself. The Access Point used for the experiments is a Linksys WRT3200ACM, operated with the latest OpenWrt version and with the CSI Sniffer tool properly installed and configured. 
The two use cases, which may have clear implications in forensics activities are the following: 
\begin{itemize}
    \item Use Case 1: Presence detection of a person moving inside a room.
    \item Use Case 2: Detecting the passage of a person through the room door.
\end{itemize}
The two different use cases are set up in the same room, with the devices placed into two different layouts. Figure \ref{fig:scenarios} sketches the different positions of the devices inside the room as well as the movements of the person involved in the experiments.



In the first use case, the transmitter (smart camera) and the receiver (CSI Sniffer) are placed on the two opposite sides of the room, while the person alternates moments in which she is not present in the room with moments in which she moves in the dashed area. In the second use case, the two devices are placed on the two sides of the door. In this latter case, the person moves in and out of the room following the dashed line. It is worth highlighting that in both use cases the camera FoV does not overlap with the area where the person is moving. This was done to (i) decouple the camera packet generation rate by the movement of the person (to avoid that changes in movement in the FoV correspond to changes in the packet generation rate) and (ii) to provide a general scenario where the IoT camera plays the role of a generic IoT device. For the same reason, during the experiments, we turned off the camera audio to avoid additional traffic generated for the transmission of sounds produced by the person. For each use case, experiments are repeated several times, each time taking note of the ground truth information (timestamps corresponding to a person occupying the room or passing through the door). In the first use case, we collected data for two hours, alternating moments in which the person is present in the room with moments where she is outside. In total, we obtained 1200 samples for each situation, in a balanced fashion. 
For the second use case, we collected the data with the person crossing the door every 30 seconds. We totally collected 50 passages (25 to get inside and 25 going outside) and the walk duration is around 3 seconds.

\subsection{Data preprocessing}
After the data collection phase, the output from CSI Sniffer is downloaded as a .CSV file and processed. We recall that such output contains one row for each captured frame, each containing the timestamp of the frame, the sender address, and the CSI value for each subcarrier.
We rely on the methodology described in \cite{9500267} for processing the CSI samples. The approach is based only on the amplitude component of each received subcarrier, discarding the phase information. Furthermore, the methodology suggests aggregating the CSI samples of the different subcarriers in a single value to be used as a reference feature.
In detail, the methodology works according to the two following steps:
\begin{enumerate}
\item\textit{Outlier removal:} for every captured frame at timestamp $t$, and every subcarrier $i$, this process substitutes any CSI amplitude value $A^i_t$ exceeding the average value $\mu^i_{t-w_1:t}$ by a specific threshold with the previous CSI sample $A^i_{t-1}$. The threshold is set to be equal to $\lambda=3$ times the standard deviation $\sigma^i_{t-w_1:t}$. Both the average and the standard deviation are computed in a window of $w_1=5$ seconds preceding the captured frame, that is:
\begin{equation}
\overline{A}_{t}^{i}  =
\begin{cases}
A_{t}^{i} &  \frac{|A_{t}^{i} - \mu^i_{t-w_1:t} |}{\sigma^i_{t-w_1:t}} < \lambda \\
A_{t-1}^{i} & \text{otherwise}
\end{cases}
\label{eq:filter}
\end{equation}




\item\textit{Subcarrier aggregation:}
 After the outlier removal step, the CSI amplitude samples are aggregated both in time and frequency. First, captured frames are grouped into non-overlapping time windows of $w_2=3$ seconds. Then, for each time window $k$, the standard deviation $\sigma^{i}_{k}$ of the CSI samples for each subcarrier $i$ is computed. Finally, the average of all the previously computed standard deviations is obtained for each window, that is:
\begin{equation}
\label{eq:meansub}
   A^*_{k} = \mu([\sigma_{k}^{i}\ \forall i \in I])
\end{equation}

Note that $I$ in \eqref{eq:meansub} is the set of available subcarriers. In our case, since we are using 20MHz channels in the 2.4 GHz band, totally we have 64 available subcarriers with index in the range [-32,31]. Considering that usually the extreme subcarriers have null amplitude and that the subcarrier with index 0 is not used for data transmission, to compute $A^*_t$ we use only the subcarriers with index in the set I = [-28,-27, ... -1, 1, ..., 27, 28], resulting in 56 subcarriers.
\end{enumerate}

\begin{figure}[t]
	\centering
	\includegraphics[width=\columnwidth]{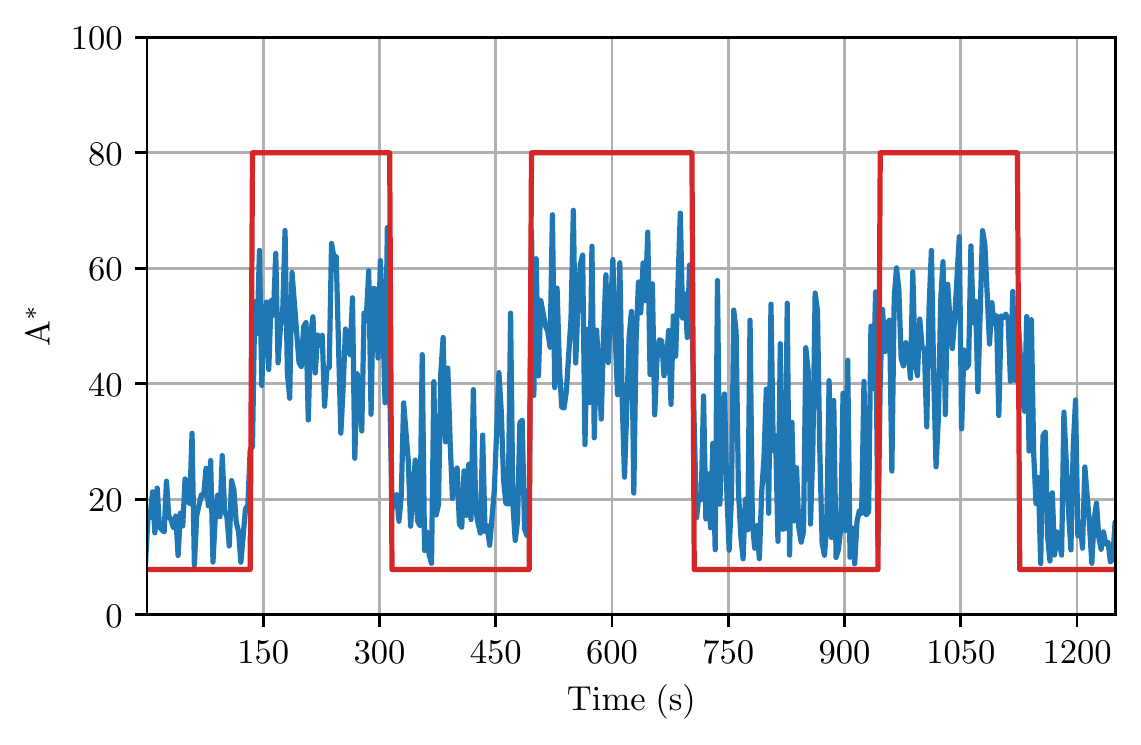}
	\caption{The value of $A^*_k$ in Use Case 1 obtained after removing outliers. The red square wave indicates when there is a person (high value) or the room is empty (low value)}
	\label{fig:Astar1}
\end{figure}

The value of $A^*_k$ is intrinsically related to the amount of movement and physical changes in the monitored environment. In a nutshell, the value of $A^*_k$ increases significantly when a person moves into the environment, while it remains low otherwise.\\
To show how the value of $A^*_k$ reacts to the presence of a person in the environment, we report its value over time in the two different considered use cases in Figures \ref{fig:Astar1} and \ref{fig:Astar2}, respectively. The red lines in the two plots refer to the ground truth, the moment in which the person is present or not for Use Case 1, and the moment in which the person crosses the door for Use Case 2, respectively.
To highlight the importance of the outliers removal phase, the figure for the second use case shows the values of $A^*_k$ when computed using the dataset before and after the filtering step. It is easy to see that the outlier removal phase removes peaks not related to the presence of the person, thus helping the classification task.

\subsection{Experimental Results}
\begin{figure}[t]
	\centering
	\includegraphics[width=\columnwidth]{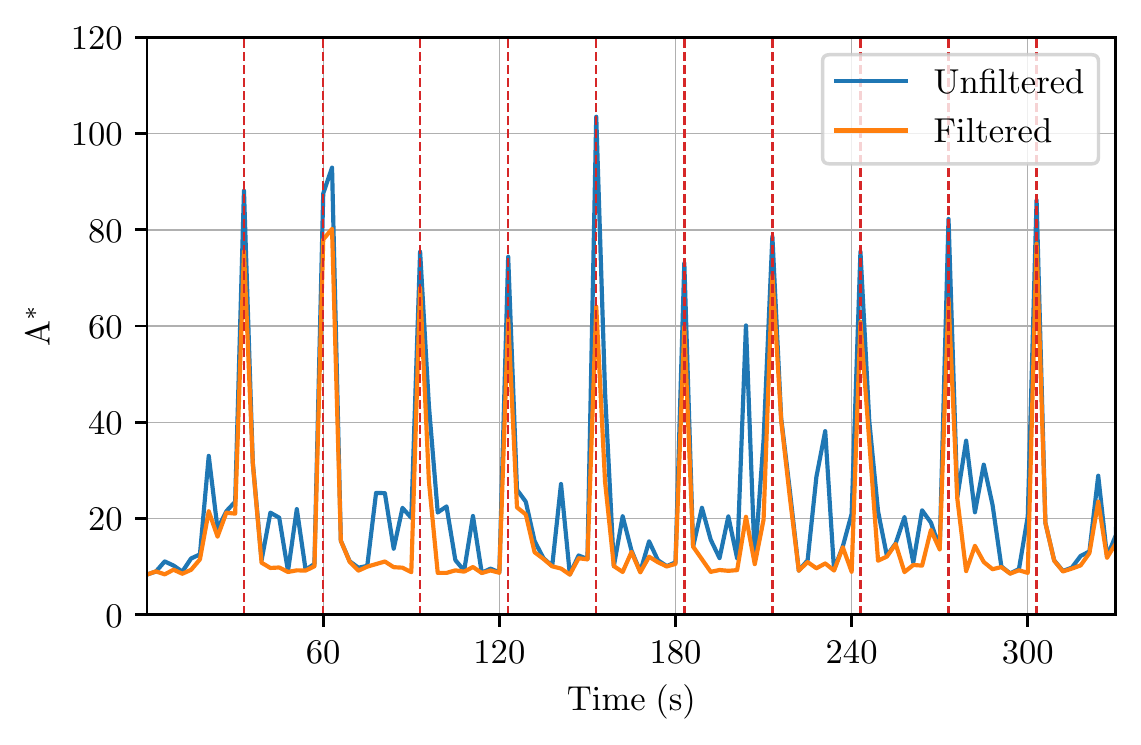}
	\caption{The value of $A^*_k$ in Use Case 2 obtained before (blue) and after (orange) removing the outliers. Red vertical lines indicate when the person is crossing the door.}
	\label{fig:Astar2}
\end{figure}

After the preprocessing steps, we obtained, for each use case, a one-dimensional time series $A^*_k$ as well as a ground-truth binary time series $G_k$ (red lines in Figures \ref{fig:Astar1} and \ref{fig:Astar2}) which takes value equal to 1 for the time windows where the human activity or a passage through the door was performed and 0 otherwise.

The next step is the design of a binary classifier, which takes as input the preprocessed time series $A^*_k$ and outputs a binary label $Y_k$ that predicts whether human activity was detected or not. We opt for a simple threshold classifier, according to the following:
\begin{equation}
Y_{k}  = 
\begin{cases}
1 & A^*_k \ge \tau \\
0 & otherwise
\end{cases}
\end{equation}
Clearly, the choice of the proper value for the threshold $\tau$ greatly impacts the obtained performance. Therefore, we performed several experiments with 1000 linearly spaced different values of $\tau$, ranging between the minimum and maximum values assumed by $A^*_k$.\\
At each iteration, we output the value of $Y_k$ for each time window and we compare it with the ground-truth label $G_k$, obtaining the number of True Positives (TP), True Negatives (TN), False Positives (FP) and False Negatives (FN).

\begin{figure}[t]
	\centering
	\includegraphics[width=\columnwidth]{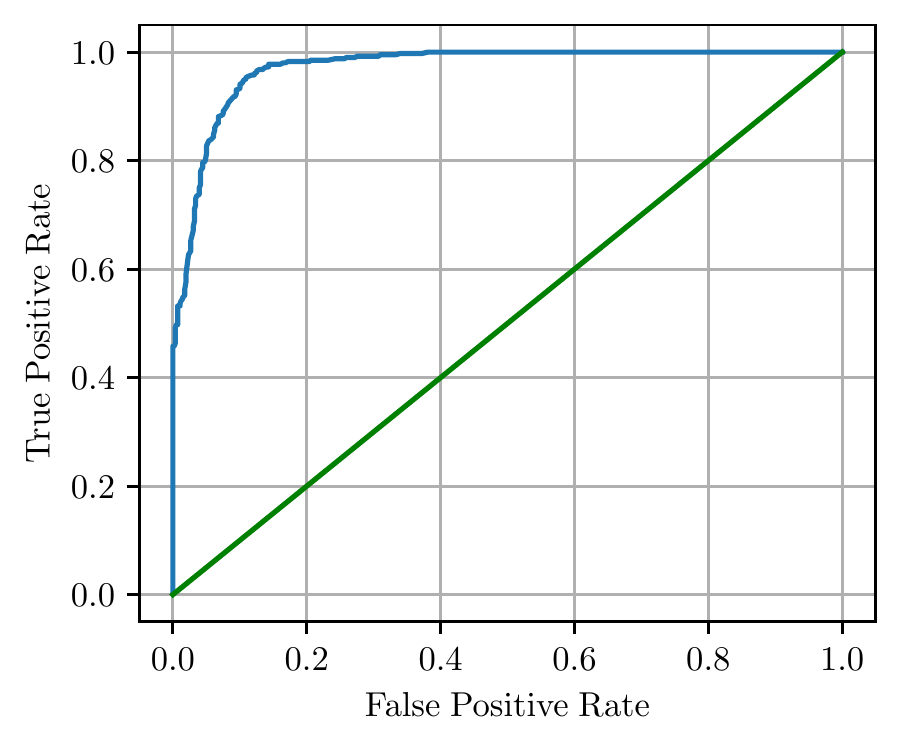}
	\caption{The ROC curve for the classification in Use Case 1. The corresponding AUC is 0.9718}
	\label{fig:roc1}
\end{figure}

\begin{figure}[t]
	\centering
	\includegraphics[width=\columnwidth]{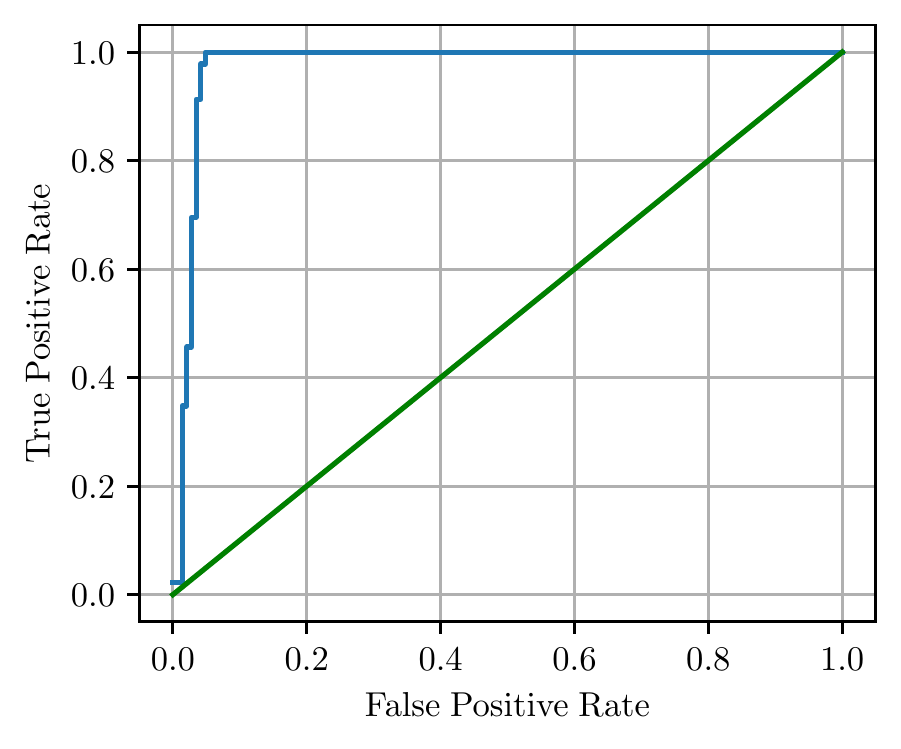}
	\caption{The ROC curve for the classification in Use Case 2. The corresponding AUC is 0.9752}
	\label{fig:roc2}
\end{figure}

From such values we extract the two following metrics:
\begin{itemize}
    \item True Positive Rate (TPR) or sensitivity: the probability that an actual positive (human activity) is classified as positive 
    \begin{align}
        TPR = \frac{TP}{TP+FN}
    \end{align}
     \item False Positive Rate (FPR) or miss rate: the probability that an actual negative (absence of human activity) is classified as positive 
    \begin{align}
        FPR = \frac{FP}{FP+TN}
    \end{align}
\end{itemize}
By representing the values of the True Positive Rate as a function of the False Positive Rate for each threshold value $\tau$ we obtain the Receiver Operating Characteristic (ROC) curve, a well-known indicator for performance measurement in binary classifications. The classification performance is better as the ROC curve is closer to the top-left point while it gets worse as the curve is closer to the diagonal.
Additionally, to have a summary indicator of the classification performance, we compute the Area Under the ROC Curve (AUC). The ideal classifier contains the point (1,1) and has an AUC of 1, while the ROC for a random classifier (e.g. throwing a coin) coincides with the diagonal and has an AUC of 0.5.\\
Figures \ref{fig:roc1} and \ref{fig:roc2} report the ROC curve computed for the two use cases, respectively. The results in terms of AUC are similar and very good in both cases, with a slightly better outcome in Use Case 2 where the AUC is 0.9752, against the 0.9718 AUC obtained in Use Case 1. These experiments demonstrate how opportunistically capturing, storing, and analyzing CSI samples from IoT traffic may pave the way for forensic analysis that may reveal important aspects of human lives. Note that, due to the higher number of samples considered in Use Case 1, the corresponding ROC curve is smoother than the one obtained for Use Case 2.

\subsection{Transmission Rate}
\begin{figure}[t]
	\centering
	\includegraphics[width=\columnwidth]{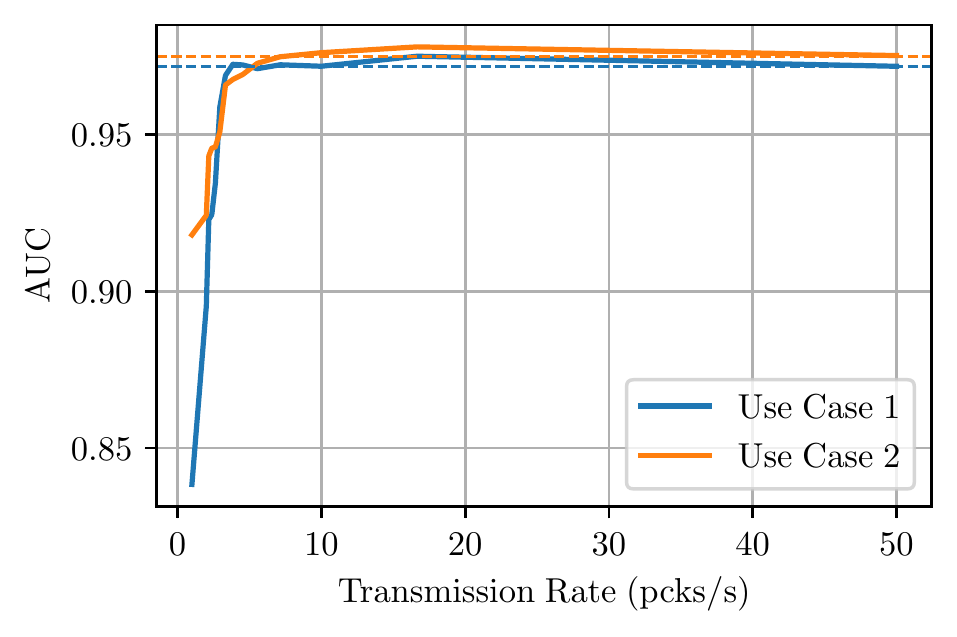}
	\caption{Classification results for the two use cases under different transmission rates}
	\label{fig:discard}
\end{figure}

In the previously presented experiments, we used an IoT Smart Camera as source of traffic. This was done primarily due to the generally high transmission frame rate of such a class of devices. Indeed, in both use cases, the average frame rate output by the Smart Camera was in the order of 30-50 frames per second.
However, it is not rare to encounter commercially available IoT devices with much lower transmission rates. As an example, smart bulbs, smart plugs and other IoT devices characterized by very simple functionalities have transmission patterns that are very different from the ones of a smart camera. 
Therefore, to generalize our analysis to devices characterized by lower transmission rates, we synthetically reduced the dataset size by dropping a specific portion of the captured frames (and the corresponding CSI samples). In detail, we produced different datasets by retaining only one every $f$ frames, with $f$ in the set $\{1\ldots25,50\}$.\\
Figure \ref{fig:discard} reports the results obtained when applying the classification using different transmission rates for the two use cases. The results are reported starting from a rate of 1 packet/s which is the minimum rate observed to achieve consistent results, up to 50 packets/s which on average is the standard transmission rate of the IoT camera used in the experiments. As the figure shows, we can achieve acceptable performance even if considering a very low number of frames. In both use cases indeed, we achieve over 90\% AUC using a device producing on average only 3 packets/s.

\begin{figure*}[t]
	\centering
	\includegraphics[width=1.8\columnwidth]{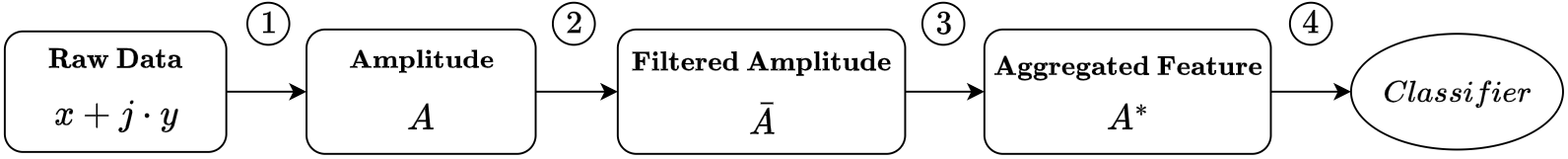}
	\caption{Pipeline of the data}
	\label{fig:data_pipeline}
\end{figure*}

\begin{figure}[t]
	\centering
	\includegraphics[width=\columnwidth]{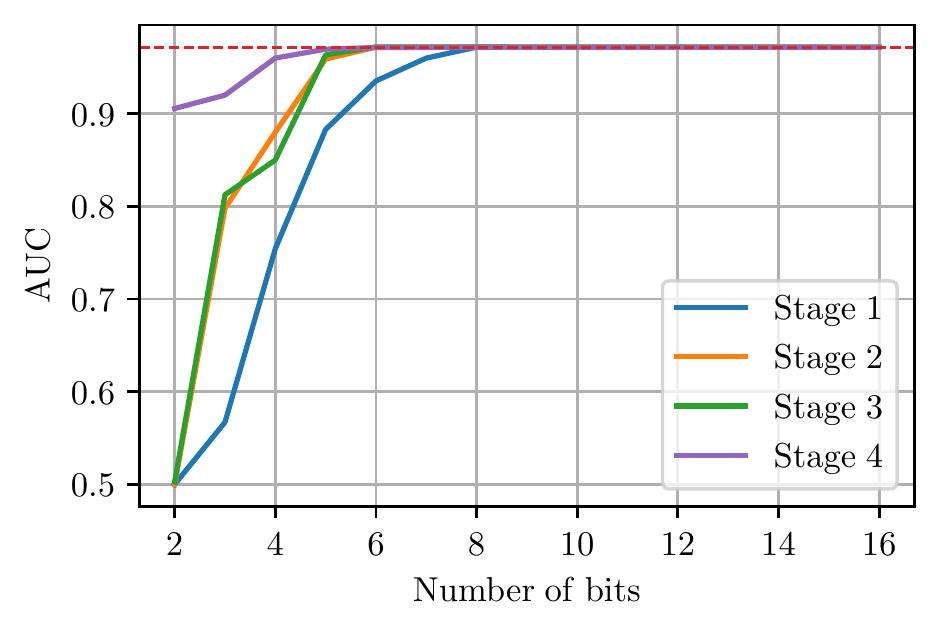}
	\caption{Classification results for Use Case 1 applying scalar quantization to the data at different stages of the pipeline (see Figure \ref{fig:data_pipeline}), using different numbers of bits}
	\label{fig:compression1}
\end{figure}

\section{Storage optimization}
\label{sec:storage}
The previous section highlighted that with the use of CSI Sniffer it is possible to perform forensic analysis tasks starting from IoT devices already present in the environment. Such forensic tasks are however commonly performed a-posteriori, requiring the data to be stored and maintained for a consistent amount of time before it is actually processed. For this reason, it is crucial to pay attention to the storage occupation of the output produced, especially considering the increasing number of IoT devices simultaneously connected in a modern smart home network. This section proposes possible solutions for optimizing the storage occupation of the CSI dataset without impacting on the forensic capabilities.

\subsection{Discarding frames}
The first trivial option is to avoid storing some of the captured frames. Indeed, the analysis presented in Figure \ref{fig:discard} shows that frame discarding is a simple solution to reduce the amount of data to be stored. Indeed, as an example for both the tested use cases, by keeping just 1 packet over 10 it is possible to save 90\% of the required storage without impacting the classification performance.


\subsection{CSI sample quantization}
One of the most common techniques for reducing a dataset size is to compress its entries before storage. In this work we opt for lossy compression algorithms, which are able to greatly reduce the required storage at the cost of introducing noise in the compressed samples. When performing forensic analysis on compressed samples, the higher the compression ratio, the higher the reconstruction noise and, as a consequence, the lower the task classification performance. 
To understand the tradeoff between the compression gain and the classification accuracy, we used scalar quantization to compress the values in the dataset. Scalar quantization allows representing each value with a fixed number of bits $B$, by first normalizing the data and then encoding the single entries as an integer in the interval $[0,2^B - 1]$, that is:



\begin{equation}
v_{i,j} = \lfloor \frac{v_{i,j}-v_{\text{min}}}{v_{\text{max}}-v_{\text{min}}} \cdot (2^{B}-1) \rceil
\label{eq:quant}
\end{equation}
The variables $v_{min}$ and $v_{max}$ in equation \ref{eq:quant} represent the minimum and maximum value that $v$ assumes in the dataset, respectively.

\begin{figure}[t]
	\centering
	\includegraphics[width=\columnwidth]{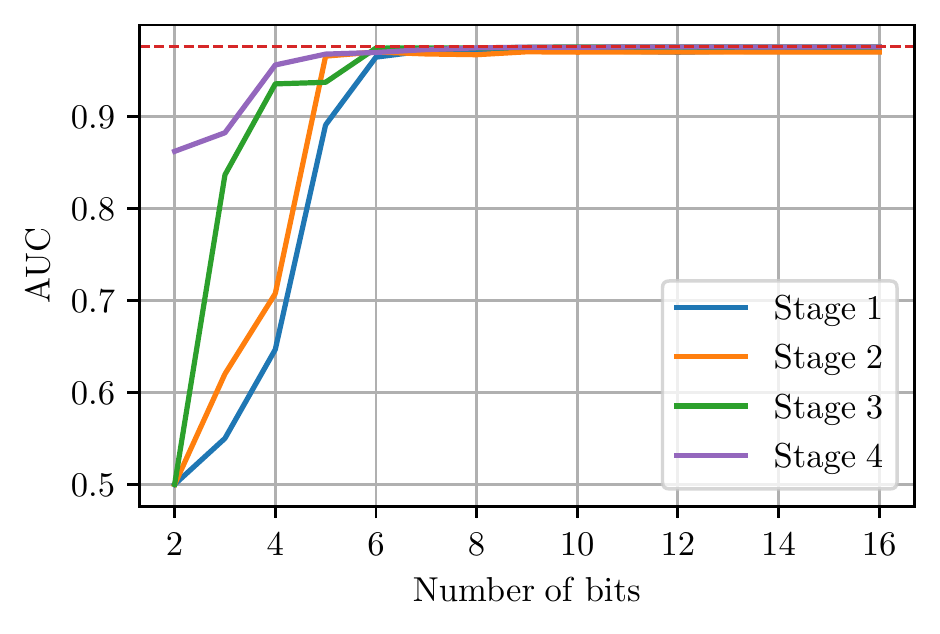}
	\caption{Classification results for Use Case 2 applying scalar quantization to the data at different stages of the pipeline (see Figure \ref{fig:data_pipeline}), using different numbers of bits}
	\label{fig:compression2}
\end{figure}

Given the data pipeline considered in this work (as sketched in Figure \ref{fig:data_pipeline}), scalar quantization can be placed in four different places:
\begin{enumerate}
\item Compress and store the raw complex numbers $(x + j\cdot y)$ for each subcarrier
    \item Compress and store the amplitude $A$ for each subcarrier
    \item Compress and store the filtered amplitude $\bar{A}$ for each subcarrier
    \item Compress and store the aggregated feature $A^*$
\end{enumerate}
The choice of the specific option also depends on the application scenario: as an example, one may want to store directly the raw complex numbers in order to be able to recover both amplitude and phase information for each subcarrier at analysis time.
We proceed thus analyzing the two use cases by applying data compression in the 4 different stages. After applying the compression in one stage, we performed the successive phases in the pipeline and eventually apply the classifier to compute the performance. For every stage, we apply scalar quantization on each dataset entry using different numbers of bits in the range (2,16), obtaining 15 new datasets (for a total of $4 \times 15 = 60$  new datasets). For each obtained dataset we compute the AUC as explained before. The results are reported in Figures \ref{fig:compression1} and \ref{fig:compression2} for the two use cases. As expected, it is clear from the figure that the best results are achieved when using data obtained with the compression of the aggregate feature involved in the classification, that is $A^*$ at Stage 4.
However, in the data pipeline, the aggregate feature is the result of all four steps and needs non-trivial processing capabilities to be computed on-the-fly, to be subsequently stored. On the other hand, the worst results are obtained when compressing the raw complex data components (real and imaginary) and operating the full pipeline with compressed data. This last case is still the most immediate to be implemented, by compressing the data as they are captured without particular extra processing.\\
In both use cases, applying quantization at stages 2-4, we only need 5 bits to achieve optimal classification results. Applying quantization to raw data instead, at least 8 bits are needed for achieving optimal results. Overall, the results are good considering that each original data entry is stored as a 32-bit floating point value for the aggregate value, and as two 16-bit integer values for raw data.

\section{Related Work}
\label{sec:related}
Wi-Fi-based human activity recognition has gained a lot of interest in scientific research in the last decade. Many research works have widely explored the potential of radio-based sensing for unveiling activities in the environment. 
The authors of \cite{csi-overview} present a detailed survey on human sensing based on Wi-Fi devices for the development of smart home systems. The work discusses many works using CSI extracted from Wi-Fi devices for daily activities recognition \cite{carm}, gesture recognition \cite{WiFinger}, and other human movement detection (e.g. lip motion for word recognition in \cite{WiHear}). Closer to the concept of IoT forensics, the authors of \cite{soto2022survey} present a recent survey on vital signs monitoring through Wi-Fi CSI. The work mainly focuses on heart rate and respiration rate monitoring which can be of vital importance in a forensic use case.
The work in \cite{7458186} presents instead a technique for detecting falls in a closed environment through the use of CSI data, processed with Machine Learning techniques. However, most of the research works present in the literature require task-specific hardware to be used under extremely controlled environments, for example with transmitter and receiver located in specific positions and devices transmitting at constant rates.\\
The authors of the ESP32 CSI tool \cite{esp32-csi} presented several works in which Wi-Fi CSI extracted using the ESP32 micro-controller as both transmitter and receiver can be used for unveiling information from the room in which the devices are located. For example, in \cite{9500267} they use CSI for detecting the presence of targets as well as their moving direction in a hallway environment with through-wall sensing, with one of the two devices located outside the interested room. 
At the same time, the data-gathering pipeline in such works is strictly related to specific hardware setups based on the ESP32 microcontroller. Our paper, on the contrary, does not assume any specific tx/rx setup and allows to extract CSI data from already deployed IoT devices and the access points managing them. Finally, for what concerns public frameworks for IoT forensics, the number of publicly available software for the smart home is very limited. The authors of \cite{fsaiot} presented a centralized controller able to simultaneously perform state acquisition and control the IoT devices.
In our previous work \cite{feature-sniffer}, we presented a framework to be installed in Wi-Fi Access Points, allowing on-the-fly extraction of network traffic features obtained aggregating packets overhead information (e.g. interarrival time, packet size), and we show possible use cases for IoT forensic analysis. In \cite{globecom-storage-accuracy}, we analyzed the storage-accuracy tradeoff of IoT forensic analysis, applying compression methodologies similar to the one we applied in this work to network traffic features.

\section{Conclusion}
\label{sec:conclusion}
This work presented \textit{CSI Sniffer}, a tool that integrates the collection of CSI features in OpenWrt-based Wi-Fi Access Points. 
After an accurate description of the tool logic, as well as its implementation details, we present two possible use cases in which the output of the tool can be used for IoT forensics tasks. We show that CSI characteristics extracted from a generic IoT device can help to unveil the presence of a person inside a room and can be used for detecting its passage through the door, without setting up additional hardware. The work concludes with an analysis of the storage/accuracy trade-off, proposing different possible solutions for optimizing the storage required for performing the defined tasks.\\
As future research directions, we plan to investigate the potential of the CSI extracted from the 5GHz band Wi-Fi channels. We are also planning to involve more IoT devices simultaneously as well as considering multiple CSI extractors in the same environment \cite{multiap-csi}. Furthermore, we will investigate advanced compression techniques for CSI samples, e.g., by selecting only a subset of the available subcarriers or leveraging the temporal correlation among them.

\bibliographystyle{IEEEtran}
\bibliography{bibliography}

\end{document}